\documentclass{ws-mpla}
\begin{document}
\title{Dark energy and the future fate of the Universe}
\author{Yungui Gong}
\address{College of Electronic Engineering, Chongqing
University of Posts and Telecommunications, Chongqing 400065,
China\\
gongyg@cqupt.edu.cn}
\author{Yuan-Zhong Zhang}
\address{ CCAST (World Laboratory), P.O. Box 8730, Beijing 100080 and\\
Institute of Theoretical Physics, Chinese Academy of Sciences,
P.O. Box 2735, Beijing 100080, China}
\maketitle
\begin{abstract}
We consider the possibility of observing the onset of the late time inflation of our
patch of the Universe. The Hubble size criterion and the event horizon criterion are applied
to several dark energy models to discuss the problem of future inflation of the Universe.
We find that the acceleration has not lasted long enough to confirm the onset of inflation by present observations
for the dark energy model with constant equation of state, the holographic dark energy model
and the generalized Chaplygin gas (GCG) model. For the flat $\Lambda$CDM model with $\Omega_{m0}=0.3$,
we find that if we use the Hubble size criterion, we need to wait until the $a_v$
which is the scale factor at the time when the onset of inflation is observed
reaches $3.59$ times of the scale factor $a_T$ when the Universe started acceleration,
and we need to wait until $a_v=2.3 a_T$ to see the onset of inflation
if we use the event horizon criterion. For the flat holographic dark energy model with $d=1$,
we find that $a_v=3.46 a_T$ with the Hubble horizon and $a_v=2.34 a_T$
with the event horizon, respectively. For the flat GCG model with the best
supernova fitting parameter $\alpha=1.2$, we find that $a_v=5.50 a_T$ with the Hubble horizon
 and $a_v=2.08 a_T$ with the event horizon, respectively.
\end{abstract}
\ccode{PACS Nos.: 04.20.Gz, 98.80.Cq}

\section{Introduction}
There are increasing evidences that the Universe is expanding with acceleration and the transition
from decelerated expansion to accelerating expansion happened in the recent past.
The transition redshift $z_T>0.2$, we use the subscript $T$ to denote the transition throughout this paper.
These results suggest that there exists dark energy (DE) with negative pressure in the Universe, and
the DE was subdominant in the past and dominates the Universe now. The presence
of DE has a lot of interesting physical effects. For example,
there exists an event horizon if the acceleration is eternal. The event
horizon sets a causal limit that the observers can ever access. The existence
of eternal acceleration also prevents us from ever measuring inflationary perturbations
which originated before the ones currently observable \cite{nbound,nbound1}. The
DE physics is still a challenging topic. The current supernova Ia (SN Ia)
data is unable to distinguish different DE models and different
DE parameterizations \cite{riess,alam,alam1,daly,daly1,hannestad,hkjbp,hkjbp1,hkjbp2,lazkoz,gong1,gong2,gong05,gong06,gong3,rcf}.
Starkman, Trodden and Vachaspati (STV)
addressed the problem of inflation in our patch of the Universe with the help of the concept of the minimal
anti-trapped surface (MAS) \cite{stv}. They argued that if we can confirm the acceleration up to a
redshift $z_c$ and observe the contraction of our MAS, then we are certain that our universe is inflating.
If we see the contraction of our MAS, then we observe the onset of inflation. The immediate conclusion is that
our universe is undergoing inflation because the cosmic acceleration is confirmed up to the redshift $1.755$
by the SN 1997ff.
STV found that the period of acceleration has not lasted long
enough for observations to confirm the onset of inflation for the $\Lambda$CDM model.

The work of STV is based on the earlier work of Vachaspati and Trodden, who
proved that in a homogeneous and isotropic universe, the necessary and
sufficient condition for observing the contraction of the MAS is that
the Universe is vacuum dominated in a region of radius greater than the Hubble
size $H^{-1}$ \cite{vt}. The comoving contraction of our MAS is the essence of inflation. Thus
only if a region of size greater than $H^{-1}$ remains vacuum dominated long enough
for the MAS to begin collapsing then we are certain that the Universe is undergoing inflation \cite{stv}.
 Because
the Hubble size is increasing with time in general, so the later the transition time
(the smaller the redshift $z_T$) is,
the longer inflation needs to last (the larger $a_v/a_T$ is, where $a_v$ is the scale factor
at the time when the onset of acceleration is first seen.).
Avelino, de Carvalho and Martins then replaced
the Hubble size by the event horizon with some additional assumptions \cite{acm}.
If the event horizon criterion is used, then the smaller $z_T$ is,
the smaller $a_v/a_T$ we need.
Huterer, Starkman and Trodden also analyzed general DE models and
found that current observations are unable to confirm the onset of inflation \cite{hst}.
In this paper, we discuss the holographic DE model \cite{hldark1,hsu,li,huang,gonghl,gongh2}
and the generalized Chaplygin gas (GCG) model \cite{chap,chap1,chap2,gonggcg}.

\section{The Hubble Size Criterion}
Vachaspati and Trodden proved that inflationary models based on the classical
Einstein equations, the weak energy conditions, and trivial topology, must assume
homogeneity on super-Hubble scales. Based on this result, STV
introduced the concept of the MAS to discuss the observability
of the onset of inflation. The MAS is a sphere, centered on the observer, on which the velocity of
comoving objects is the speed of light $c$ \cite{stv}. For light emitted directly toward the observer
inside the MAS, the photons get closer to the observer with time, while all photons emitted by sources
outside the MAS get farther away. For a homogeneous and isotropic universe, the physical radius of the MAS
at time $t$ is the Hubble size $1/H(t)$. It was argued that the beginning of the comoving contraction
of the MAS can be identified with the onset of inflation. Note that the condition of the onset of inflation is
$\ddot{a}=0$ which is equivalent to $d(aH)^{-1}/dt$, where $a(t)$ is the scale factor.

If a light was emitted at time $t_e$ from a source located
at a comoving distance $r$ and then is received at time $t_v$ by the observer located at the origin,
the physical distance between the source and the observer at $t_e$ is
\begin{equation}
\label{phydis}
d(t_e,t_v)=a(t_e)\int^{t_v}_{t_e}\frac{dt}{a(t)}.
\end{equation}
In this paper, we consider a flat universe. We can see the
contraction of the MAS at time $t_v$ if $d(t_e,t_v)=1/H(t_e)$ \cite{stv}, here $t_v$ ($a_v$)
is the time (scale factor) when the turnaround of the MAS comes into view. Therefore,
the onset of the late time inflation can be seen at time $t_v$ if $d(t_T,t_v)=1/H(t_T)$, where $t_T$
is the transition time when the Universe experienced from the deceleration phase to the acceleration phase.
To see the consequences of $d(t_e,t_v)=1/H(t_e)$ clearly, we use a simple DE model
$p=w\rho$ with constant $w$ satisfying $w\ge -1$ as an example.

\begin{figure}
\centering
\includegraphics[width=12cm]{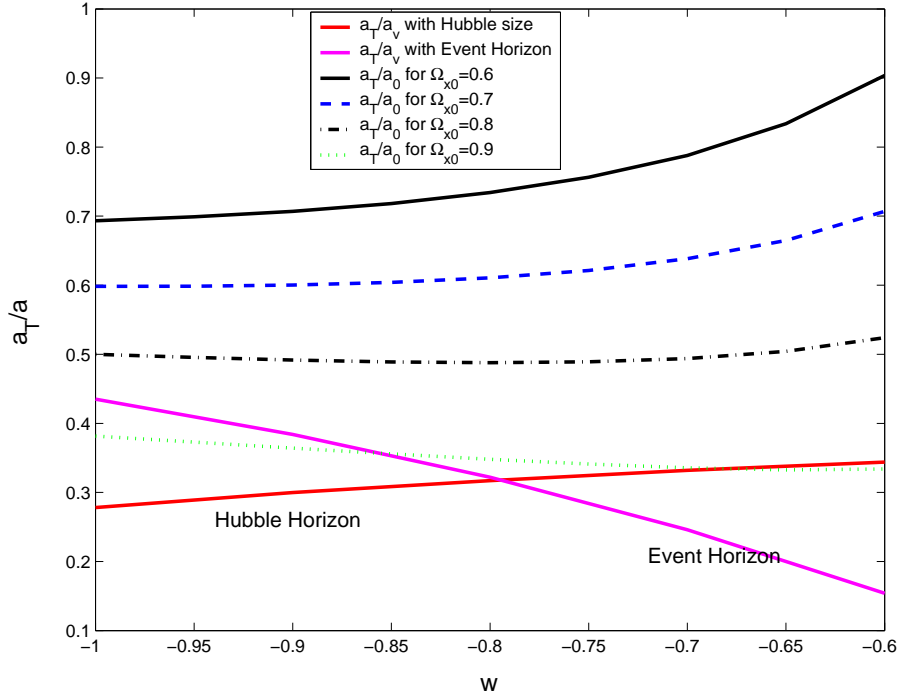}
\caption{The dependence of $a_T/a_0$ and $a_T/a_v$ on $w$ and $\Omega_{x0}$ for
the DE model with constant equation of state parameter $w$. The line labeled ``Hubble Horizon" shows the dependence
of $a_T/a_v$ on $w$ by using the Hubble scale criterion. The line labeled ``Event Horizon" shows the dependence
of $a_T/a_v$ on $w$ by using the event horizon criterion.} \label{fig1}
\end{figure}

The matter energy density is $\rho_m=\rho_{mr}(a_r/a)^3$, where the subscript $r$ means that
the variable takes a value at an arbitrary reference time $t_r$. The DE density is
$\rho_x=\rho_{xr}(a_r/a)^{3(1+w)}$. So we have
\begin{equation}
\label{hubeq1}
H^2=H^2_r\left[\Omega_{mr}\left(\frac{a_r}{a}\right)^3+\Omega_{xr}\left(\frac{a_r}{a}\right)^{3(1+w)}\right],
\end{equation}
where $\Omega=8\pi G\rho/(3H^2)$.
The transition time $t_T$ is determined from
\begin{equation}
\label{acc1}
\Omega_{mT}+(1+3w)\Omega_{xT}=0,
\end{equation}
or
\begin{equation}
\label{acc2}
1+z_T=\frac{a_0}{a_T}=\left(-\frac{\Omega_{m0}}{\Omega_{x0}(1+3w)}\right)^{1/3w},
\end{equation}
where $\Omega_{m0}=1-\Omega_{x0}$ and the subscript $0$ means that
the variable takes its present value. For the $\Lambda$CDM model, $w=-1$ and
$\Omega_{m0}=0.3$, we get $z_T=0.67$.
Using Eq. (\ref{acc1}), the condition $d(t_T,t_v)=1/H(t_T)$ becomes
\begin{equation}
\label{wcond1}
\int^1_{a_T/a_v}\frac{dx}{\sqrt{x^3(x^{3w}-1-3w)}}=\frac{1}{\sqrt{-3w}}.
\end{equation}
To be able to observe the onset of inflation at present, we require $a_v<a_0$. For the $\Lambda$CDM model,
the solution to the above equation (\ref{wcond1}) is $a_v/a_T=3.59>1+z_T=1.67$, so $a_v>a_0$
and $\Omega_{\Lambda v}=0.96$.
In addition, we require the confirmation of cosmic acceleration up to a redshift $z_c$,
where $z_c$ determined from the condition
$d(t_c,t_0)=1/H(t_c)$ satisfies the following equation
\begin{equation}
\label{wcond2}
\int^{z_c}_0\frac{dz}{\sqrt{\Omega_{m0}(1+z)^3+\Omega_{x0}(1+z)^{3(1+w)}}}=
\frac{1+z_c}{\sqrt{\Omega_{m0}(1+z_c)^3+\Omega_{x0}(1+z_c)^{3(1+w)}}}.
\end{equation}
Note that $z_c$ is the minimum redshift that
the cosmic acceleration needs to be observed by current observations.
For the $\Lambda$CDM model with $\Omega_{m0}=0.3$, we get $z_c=1.61$. Since the current
SN Ia observations extend to redshift $1.755$, so the cosmic acceleration is confirmed. But we still
need to wait until $\Omega_\Lambda$ reaches a value of $0.96$ to observe the onset of inflation.
The solutions to Eqs. (\ref{acc2})
and (\ref{wcond1}) for other choices of $\Omega_{x0}$ and $w$ are shown in Fig \ref{fig1}. From Fig.
\ref{fig1}, we see that we need to wait until $\Omega_{x}\sim 0.9$ for $w > -0.67$ to be able
to observe the onset of inflation. For bigger $w$, we need smaller $\Omega_x$. However, current
observations strongly constrain $w\lesssim -0.8$. In other words, we are unable to confirm the onset of inflation now
with current observations.

In Ref.~\refcite{press}, Lightman and Press introduced the concept of constant redshift surfaces to discuss
the causal communication of comoving particles. In the definition of the constant redshift surfaces, the lower
integral $t_e$
changes with the upper integral $t_v$ in equation (\ref{phydis}) by fixing $a(t_v)/a(t_e)=z_1$ to be a constant.
The constant redshift surface or $z_1$-surface increases with time before inflation. After inflation,
the $z_1$-surface will eventually decrease with time. For small redshift $z_1<z_T$, the $z_1$-surface
is decreasing. For large redshift, the $z_1$-surface is increasing. So there exists a turnaround
redshift $z_1$ so that the $z_1$-surface reaches its maximum at present. For the $\Lambda$CDM model,
we find that the turnaround redshift $z_1=2.09$ if we take $\Omega_{m0}=0.3$. Although the
decrease of the $z_1$-surface for small $z_1$ is a characteristic feature of an accelerated universe,
it does not mean that the Universe is inflating and the space-time will evolve into
the de-Sitter phase. Only if $z_1$ is big enough so that the $z_1$-surface crosses the event horizon, then
we can say that the Universe is inflating. We will discuss this in the next section.

\subsection{Holographic DE model}
Cohen, Kaplan and Nelson proposed that for any state with energy $E$ in the Hilbert space,
the corresponding Schwarzschild radius $R_S\sim E$ is less
than than the infrared (IR) cutoff $L$ \cite{hldark1}.
Therefore, the maximum entropy is $S_{BH}^{3/4}$. Under this assumption, a relationship
between the ultraviolet cutoff $\rho_x^{1/4}$ and the IR cutoff is derived, i.e.,
$8\pi G L^3 \rho_x/3\sim L$. Hsu found that the model based on the Hubble scale as the
IR cutoff would not give an accelerating universe \cite{hsu}. Li then showed that a plausible dark energy
is possible by choosing the future event horizon as the IR cutoff \cite{li}. So
the holographic DE density is \cite{li,huang}
\begin{equation}
\label{holorho}
\rho_x=\frac{3d^2}{8\pi G R_{eh}^2},
\end{equation}
where $R_{eh}(t)=d(t,\infty)$ is the event horizon.
The equation of state of the holographic DE is
$$w_x=-\frac{1}{3}\left(1+\frac{2\sqrt{\Omega_x}}{d}\right).$$
Because of some physical constraints
on $d$, we take $d=1$ for simplicity \cite{gonghl}. Note that the weak energy condition is
satisfied as long as $d^2\ge \Omega_x$. By using the Friedmann equations,
we get
\begin{equation}
\label{holorho2}
\frac{d\Omega_x}{d\ln a}=\Omega_x(1-\Omega_x)(1+2\sqrt{\Omega_x}).
\end{equation}
The solution to Eq. (\ref{holorho2}) is
\begin{equation}
\label{holorho3}
\ln\Omega_x-\frac{1}{3}\ln(1-\sqrt{\Omega_x})+\ln(1+\sqrt{\Omega_x})-\frac{8}{3}\ln(1+2\sqrt{\Omega_x})
=\ln\left(\frac{a}{a_r}\right)+y_r,
\end{equation}
where $y_r$ is determined from the above equation by using $\Omega_{xr}$.
From the definition of the holographic DE density (\ref{holorho}) and
the Friedmann equations, we get
\begin{equation}
\label{ehor}
R_{eh}(t)=\frac{1}{H_r\sqrt{1-\Omega_{xr}}}\left(\frac{a(t)}{a_r}\right)^{3/2}\left(
\frac{1-\Omega_x(t)}{\Omega_x(t)}\right)^{1/2}.
\end{equation}
The transition time $t_T$ is
determined from
\begin{equation}
\label{hacc1}
\Omega_{xT}+2\Omega_{xT}\sqrt{\Omega_{xT}}=1.
\end{equation}
So $\Omega_{xT}=0.432$ and $y_T=-2.215$. Substitute these values to Eq. (\ref{holorho3}), set
$a_r=a_0$ and $\Omega_x=\Omega_{xT}$  and use the best
fitting value $\Omega_{xr}=\Omega_{x0}=0.75$, we get $z_T=0.72$.
Since $d(t_T,t_v)=R_{eh}(t_T)-a_T R_{eh}(t_v)/a_v$, the condition $d(t_T,t_v)=1/H(t_T)$ becomes
\begin{equation}
\label{hcond1}
\frac{a_v(1-\Omega_{xv})}{a_T \Omega_{xv}}=\frac{(1-\sqrt{\Omega_{xT}})^2(1-\Omega_{xT})}{\Omega_{xT}}=0.154.
\end{equation}
Combining Eqs. (\ref{holorho3}) and (\ref{hcond1}), we get
\begin{equation}
\label{omxv}
\ln\Omega_{xv}-\frac{1}{3}\ln(1-\sqrt{\Omega_{xv}})+\ln(1+\sqrt{\Omega_{xv}})-
\frac{8}{3}\ln(1+2\sqrt{\Omega_{xv}})
=\ln\left(\frac{0.154\Omega_{xv}}{1-\Omega_{xv}}\right)+y_T.
\end{equation}
Combining Eqs. (\ref{hcond1}) and (\ref{omxv}), we get $a_v/a_T=3.46>1+z_T=1.72$,
so $a_v>a_0$.
The redshift $z_c$ satisfies the equation
\begin{equation}
\label{hcond2}
\frac{(1-\sqrt{\Omega_{xc}})^2(1-\Omega_{xc})}{(1+z_c)\Omega_{xc}}
=\frac{1-\Omega_{x0}}{\Omega_{x0}}.
\end{equation}
Combining Eqs. (\ref{holorho3}) and (\ref{hcond2}) with $\Omega_{x0}=0.75$,
we get $z_c=1.64<1.755$. Therefore although we see our MAS today, we are unable to observe the
onset of inflation for the holographic DE model because the cosmic acceleration has not lasted
long enough.

\subsection{GCG Model}
\begin{figure}
\centering
\includegraphics[width=12cm]{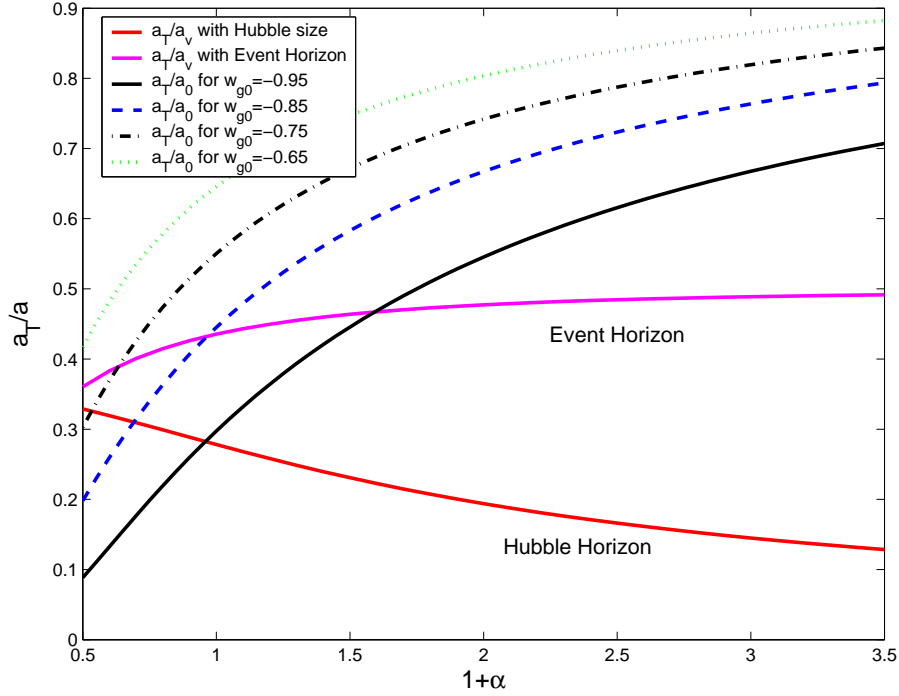}
\caption{The dependence of $a_T/a_0$ and $a_T/a_v$ on $w_{g0}$ and $\alpha$ for the GCG model.
The line labeled ``Hubble Horizon" shows the dependence
of $a_T/a_v$ on $\alpha$ by using the Hubble scale criterion. The line labeled ``Event Horizon" shows the dependence
of $a_T/a_v$ on $\alpha$ by using the event horizon criterion.} \label{fig2}
\end{figure}

For the GCG model, we have $p_g=-A/\rho_g^\alpha$. By using the energy conservation equation, we get
\begin{equation}
\label{grho1}
\rho_g=\rho_{gr}\left[-w_{gr}+(1+w_{gr})\left(\frac{a_r}{a}\right)^{3(1+\alpha)}\right]^{1/(1+\alpha)},
\end{equation}
where the equation of state parameter $w_g=p_g/\rho_g$. Because $\rho_g\sim (a_0/a)^3$
when $a\ll a_0$ and $\rho_g\sim {\rm constant}$ when
$a\gg a_0$, the GCG model can be thought as a unified model
of DE and dark matter. Therefore we assume that there is no matter present for simplicity and require
$w_g\ge -1$ so that the weak energy condition is satisfied. As discussed in Ref. \refcite{gonggcg},
some reasonable physical constraints also require $\alpha\ge 0$. The Friedmann equation is
\begin{equation}
\label{gcos1}
H^2=H^2_r\left[-w_{gr}+(1+w_{gr})\left(\frac{a_r}{a}\right)^{3(1+\alpha)}\right]^{1/(1+\alpha)}.
\end{equation}
At the transition time $t_T$, we have $w_{gT}=-1/3$. So the transition redshift satisfies
\begin{equation}
\label{gzt}
1+z_T=\left(-\frac{2w_{g0}}{1+w_{g0}}\right)^{1/3(1+\alpha)}.
\end{equation}
By using the best supernova fitting values $w_{g0}=-0.83$ and
$\alpha=1.20$, we get $z_T=0.412$. The condition $d(t_T,t_v)=1/H(t_T)$ gives
\begin{equation}
\label{gcond1}
\int^1_{a_T/a_v}\left[\frac{1}{3}+\frac{2}{3}x^{3(1+\alpha)}\right]^{-1/2(1+\alpha)} dx=1.
\end{equation}
If we take $\alpha=1.2$, we get $a_v/a_T=5.50$ and $a_v>a_0$.
The condition $d(t_c,t_0)=1/H(t_c)$ gives
\begin{eqnarray}
\label{gcond2}
\int^{z_c}_0\left[-w_{g0}+(1+w_{g0})(1+z)^{3(1+\alpha)}\right]^{-1/2(1+\alpha)} dz= \nonumber \\
\frac{1+z_c}{\left[-w_{g0}+(1+w_{g0})(1+z_c)^{3(1+\alpha)}\right]^{1/2(1+\alpha)}}.
\end{eqnarray}
By using the best supernova fitting values $w_{g0}=-0.83$ and
$\alpha=1.20$, we get $z_c=1.424<1.755$. Again the present observations of cosmic acceleration
extend to a redshift $z_c=1.424$. Currently we are still unable to confirm the onset of
inflation for the GCG model. For some other values of $w_{g0}$ and $\alpha$, the numerical solutions
to Eqs. (\ref{gzt}) and (\ref{gcond1}) are shown in Fig. \ref{fig2}. From Fig. \ref{fig2}, we see that it is
possible to confirm the onset of inflation for the GCG model only if $\alpha<0$ which is
outside the physical parameter space.

\section{The Event Horizon Criterion}
In general, the Hubble size increases with time. Therefore, it will take longer time to observe
the onset of inflation if the transition happened at later time. In order to avoid this situation,
Avelino, de Carvalho and Martins replaced the Hubble scale criterion discussed in the previous section
by requiring that the comoving distance equals to the comoving event horizon at the time of reception. Of course,
some additional assumptions on the content of the local universe and field dynamics are needed. The event
horizon criterion is
\begin{equation}
\label{ereq}
r=\int_{t_e}^{t_v}\frac{dt}{a(t)}=\int_{t_v}^\infty \frac{dt}{a(t)}.
\end{equation}
By using the notation of the constant redshift surface in \cite{press}, the
above condition gives us the redshift $z_1$ when the $z_1$-surface crosses the event horizon.
To illustrate the effect of this condition, we take the simple DE model with constant
equation of state as an example. Applying the condition (\ref{ereq}) to the onset of inflation,
Eq. (\ref{wcond1}) is replaced by
\begin{equation}
\label{wecond1}
\int^1_{a_T/a_v}\frac{dx}{\sqrt{x^3(x^{3w}-1-3w)}}=\int^{a_T/a_v}_0\frac{dx}{\sqrt{x^3(x^{3w}-1-3w)}},
\end{equation}
and Eq. (\ref{wcond2}) is replaced by
\begin{equation}
\label{wecond2}
\int^{z_c}_0\frac{dz}{\sqrt{\Omega_{m0}(1+z)^3+\Omega_{x0}(1+z)^{3(1+w)}}}
=\int^{0}_{-1}\frac{dz}{\sqrt{\Omega_{m0}(1+z)^3+\Omega_{x0}(1+z)^{3(1+w)}}}.
\end{equation}
For the $\Lambda$CDM model, the solution to Eq. (\ref{wecond1})
is $a_v/a_T=2.30>1+z_T=1.67$ or $\Omega_{\Lambda v}=0.86$, and the solution to Eq. (\ref{wecond2})
is $z_c=1.81$ if we take $\Omega_{m0}=0.3$.
So we are unable to confirm the onset of inflation now
with current observations. The numerical
solutions to Eq. (\ref{wecond1}) for other values of $w$ are shown in Fig \ref{fig1}. From Fig.
\ref{fig1}, we see that we need to wait until $\Omega_{x}\sim 0.9$ for $w< -0.85$ to be able
to observe the onset of inflation. The smaller $w$ is, the sooner we observe the onset of inflation.
It is possible that we observe the onset of inflation with $\Omega_{\Lambda 0}=0.7$ if $w<-1$. However,
the weak energy condition is violated if $w<-1$, and the criterions discussed in this paper do not apply.
This situation needs to be investigated more carefully and are out of the scope of our discussion.

\subsection{Holographic DE model}
Applying the event horizon criterion (\ref{ereq}) to the holographic DE model discussed in the
previous section, we replace Eq. (\ref{hcond1}) by
\begin{equation}
\label{hecond1}
\frac{a_v(1-\Omega_{xv})}{a_T \Omega_{xv}}=\frac{1-\Omega_{xT}}{4\Omega_{xT}}=0.329,
\end{equation}
and Eq. (\ref{hcond2}) by
\begin{equation}
\label{hecond2}
\frac{1-\Omega_{xc}}{4(1+z_c)\Omega_{xc}}
=\frac{1-\Omega_{x0}}{\Omega_{x0}}.
\end{equation}
Combining Eqs. (\ref{holorho3}) and (\ref{hecond1}), we get $a_v/a_T=2.34>1+z_T=1.72$,
so $a_v>a_0$.
Combining Eqs. (\ref{holorho3}) and (\ref{hecond2}) with $\Omega_{x0}=0.75$, we get $z_c=1.84$.
Therefore we are unable to observe the onset of inflation now for the holographic DE model.

\subsection{GCG Model}
Applying the event horizon criterion (\ref{ereq}) to the GCG model discussed in the
previous section, we replace Eq. (\ref{gcond1}) by
\begin{equation}
\label{gecond1}
\int^1_{a_T/a_v}\left[\frac{1}{3}+\frac{2}{3}x^{3(1+\alpha)}\right]^{-1/2(1+\alpha)} dx=
\int^{a_T/a_v}_{0}\left[\frac{1}{3}+\frac{2}{3}x^{3(1+\alpha)}\right]^{-1/2(1+\alpha)} dx,
\end{equation}
and Eq. (\ref{gcond2}) by
\begin{eqnarray}
\label{gecond2}
\int^{z_c}_0\frac{dz}{[-w_{g0}+(1+w_{g0})(1+z)^{3(1+\alpha)}]^{1/2(1+\alpha)}}=\nonumber\\
\int^0_{-1}\frac{dz}{[-w_{g0}+(1+w_{g0})(1+z)^{3(1+\alpha)}]^{1/2(1+\alpha)}}.
\end{eqnarray}
If we take $\alpha=1.2$, we get $a_v/a_T=2.08$, so $a_v>a_0$. By using $w_{g0}=-0.83$ and
$\alpha=1.20$, the solution to Eq. (\ref{gecond2}) is $z_c=1.64$. Because $a_v>a_0$,
currently we are unable to confirm the onset of
inflation for the GCG model. For different values of $\alpha$, the numerical solutions
to Eq. (\ref{gecond1}) are shown in Fig. \ref{fig2}. From Fig. \ref{fig2}, we see that it is
possible to confirm the onset of inflation for the GCG model
when $\alpha< 0.5$ and $w_{g0}\sim -0.95$. The smaller $\alpha$ is, the smaller $w_{g0}$ is required to
observe the onset of inflation.

\section{Discussion}
For the Hubble size criterion, our results are: (1) Constant $w\ge -1$ model. It is possible to observe
the onset of inflation when $\Omega_x\sim 0.9$ and $w> -0.67$. (2) The holographic
DE model with $d=1$. We find that $z_T=0.72$ and $z_c=1.64$ if we use the best
supernova fitting result $\Omega_{x0}=0.75$. We also get $a_v/a_T=3.46$ and $a_v>a_0$.
(3) The GCG model. We find that $z_T=0.412$ and $z_c=1.424$ if we use the best
supernova fitting results $w_{g0}=-0.83$ and $\alpha=1.20$. By using $\alpha=1.20$,
we get $a_v/a_T=5.50$ and $a_v>a_0$. It is possible to observe the onset
of inflation when $\alpha<0$ and $w_g\sim -0.75$.

For the event horizon criterion, our results are: (1) Constant $w\ge -1$ model. It is possible to observe
the onset of inflation when $\Omega_x\sim 0.9$ and $w< -0.85$. (2) The holographic
DE model with $d=1$. We find that $z_c=1.84$ if we use the best
supernova fitting result $\Omega_{x0}=0.75$. We also get $a_v/a_T=2.34$ and $a_v>a_0$.
(3) The GCG model. We find that $z_c=1.64$ if we use the best
supernova fitting results $w_{g0}=-0.83$ and $\alpha=1.20$.
By using $\alpha=1.20$, we get $a_v/a_T=2.08$ and $a_v>a_0$. It is possible to observe the onset
of inflation when $\alpha< 0.5$ and $w_g\sim -0.95$.

In general, the event horizon criterion gives bigger value of $z_c$ than the Hubble size criterion does.
The reason is that today we may be able to observe a larger portion of the Universe than that
we can ever access. The event horizon criterion is not applicable for very low values of the
dark energy density. However, the later the cosmic acceleration started, the sooner we are able to observe
the onset of inflation by using the event horizon criterion.

For all the three models discussed in this paper, we have $z_c>1$ and $a_v>a_0$. Therefore it is impossible
to confirm the onset of inflation by current observations for all three models.
However, this conclusion cannot be extended to phantom models because they violate the weak energy condition.
Therefore, the conclusion cannot apply to the holographic DE model with phantom behavior \cite{nojiri}.
However, it can be applied to the interacting holographic DE model discussed in Refs. ~\refcite{intde} and \refcite{intde1} because
$w_{tot}\ge -1$. The conclusion is neither applicable to the GCG model
with $w_g<-1$ discussed in Ref.~\refcite{ggcg}.
When we parameterized DE equation of state, we find that the supernova data might
not be able to distinguish those parameterizations that have almost the same past behaviors and different
future behaviors \cite{gong05}.
For general dark energy model, it is impossible that cosmic acceleration started at a
redshift $z_T>1$, so it is impossible to observe the onset of inflation up to a region
of Hubble size. If the event horizon criterion is used, then it is possible to observe
the onset of inflation and confirm the inflation of our universe for some general
dynamic dark energy models with low transition redshift.
In this paper, we find that the acceleration has not lasted long enough
for observations to confirm that we are undergoing inflation. Therefore the future fate of the
Universe is still unknown from current observations. We need to wait some time to be
confident that we are undergoing inflation. On the other hand, if the cosmic acceleration never
ends, less information about the early inflationary perturbations will be observed in
the future. So we are living in a peculiar era in the history of the Universe.

For the GCG model, the marginal allowed parameter spaces $\alpha \lesssim 0.5$ and $w_{g0}\gtrsim -0.75$ will make it
possible to confirm that our universe is inflating by using the event horizon criterion.

\section*{Acknowledgments}
Y. Gong is supported by NNSFC under grant No. 10447008 and 10605042,
SRF for ROCS, State Education Ministry, CQMEC under grant No.
KJ060502. Y.Z. Zhang's work was in part supported by NNSFC under
Grant No. 90403032 and also by National Basic Research Program of
China under Grant No. 2003CB716300.

\end{document}